\documentclass[aps,prl,reprint,twocolumn,superscriptaddress,showpacs,amssymb]{revtex4-1}
\usepackage[english]{babel}
\usepackage{graphics}
\usepackage{graphicx}
\usepackage{epsfig}
\usepackage{amssymb}
\usepackage{dcolumn}% Align table columns on decimal point
\usepackage{bm}% bold math
\usepackage{color}
\usepackage{natbib}
\usepackage{lineno}
\usepackage{sidecap}
\usepackage{verbatim}  % Needed for the "comment" environment to make LaTeX comments
\usepackage{vector}  % Allows "\bvec{}" and "\buvec{}" for "blackboard" style bold vectors in maths
\usepackage{wrapfig}
\usepackage{enumerate}
\usepackage{sidecap}
\usepackage{amsmath}
\usepackage{hyperref}

\begin{document}
% \preprint{0}

\title{Spectroscopic evidence of topological phase transition in 3D Dirac semimetal Cd$_3$(As$_{1-x}$P$_x$)$_2$}

\author{S.\ Thirupathaiah}
\email{hitiru@gmail.com}
\affiliation{Leibniz Institute for Solid State Research, IFW Dresden, D-01171 Dresden, Germany.}
\affiliation{Solid State and Structural Chemistry Unit, Indian Institute of Science, Bangalore, Karnataka, 560012, India.}
\author{I. \ Morozov}
\affiliation{Leibniz Institute for Solid State Research, IFW Dresden, D-01171 Dresden, Germany.}
\affiliation{Lomonosov Moscow State University, 119991 Moscow, Russia.}
\affiliation{Lebedev Physical Institute, Russian Academy of Sciences, 119991 Moscow, Russia.}
\author{Y.\ Kushnirenko}
\affiliation{Leibniz Institute for Solid State Research, IFW Dresden, D-01171 Dresden, Germany.}
\author{A. V.\ Fedorov}
\affiliation{Leibniz Institute for Solid State Research, IFW Dresden, D-01171 Dresden, Germany.}
\author{E.\ Haubold}
\affiliation{Leibniz Institute for Solid State Research, IFW Dresden, D-01171 Dresden, Germany.}
\author{T. K.\ Kim}
\affiliation{Diamond Light Source, Harwell Campus, Didcot, OX11 0DE, UK.}
\author{G.\ Shipunov}
\affiliation{Lomonosov Moscow State University, 119991 Moscow, Russia.}
\author{A.\ Maksutova}
\affiliation{Lomonosov Moscow State University, 119991 Moscow, Russia.}
\author{O.\ Kataeva}
\affiliation{Leibniz Institute for Solid State Research, IFW Dresden, D-01171 Dresden, Germany.}
\affiliation{A.E. Arbuzov Institute of Organic and Physical Chemistry, Federal Research Center, "Kazan Scientific Center of the Russian Academy of Sciences".}
\author{S.\ Aswartham}
\affiliation{Leibniz Institute for Solid State Research, IFW Dresden, D-01171 Dresden, Germany.}
\author{B. B\"uchner}
\affiliation{Leibniz Institute for Solid State Research, IFW Dresden, D-01171 Dresden, Germany.}
\author{S. V.\ Borisenko}
\email{s.borisenko@ifw-dresden.de}
\affiliation{Leibniz Institute for Solid State Research, IFW Dresden, D-01171 Dresden, Germany.}

\date{\today}

\begin{abstract}
    We study the low-energy electronic structure of three-dimensional Dirac semimetal, Cd$_3$(As$_{1-x}$P$_x$)$_2$ [$x$ = 0 and 0.34(3)], by employing the angle-resolved photoemission spectroscopy (ARPES). We observe that the bulk Dirac states in Cd$_3$(As$_{0.66}$P$_{0.34}$)$_2$ are gapped out with an energy of 0.23 eV, contrary to the parent Cd$_3$As$_2$ in which the gapless Dirac states have been observed. Thus,  our results confirm the earlier predicted topological phase transition in Cd$_3$As$_2$  with perturbation. We further notice that the critical P substitution concentration,  at which the two Dirac points that are spread along the $c$-axis in Cd$_3$As$_2$ form a single Dirac point at $\Gamma$,  is much lower [x$_c$(P)$<$ 0.34(3)] than the predicted value of x$_c$(P)=0.9. Therefore, our results suggest that the nontrivial band topology of Cd$_3$As$_2$ is remarkably sensitive to the P substitution and can only survive over a narrow substitution range, i.e., 0 $\leq$ x (P) $<$ 0.34(3).

\end{abstract}
%\pacs{}

\maketitle

%Materials showing spin polarized massless Dirac fermions in the vicinity of the Fermi level are of great interest for their potential applications in the electronic devices.
The three-dimensional (3D) Dirac semimetals~\cite{Young2012, Wang2012a, Wang2013a,  Borisenko2014, Neupane2014a, Liu2014a} with their novel band structure and unusual physical properties have attracted a great deal of research interest recently due to the potential applications~\cite{Nayak2008}. In contrast to the topological insulators where the Dirac fermions are either the edge or the surface states that are  protected by the time reversal symmetry (TRS) and adiabatically connecting the gapped bulk states~\cite{Hasan2010}, in the 3D Dirac-semimetals the Dirac fermions are bulk in nature and are protected by both the TRS and the crystal symmetries~\cite{Burkov2011, Young2012, Wang2012a}. Thus, the 3D Dirac semimetals have a clear advantage over the topological insulators from the applications point of view as the surface states generally suffer from the sample degradation with the time.

In so far reported 3D Dirac semimetals, Na$_3$Bi and Cd$_3$As$_2$,  multiple Dirac nodes near the Fermi level have been observed~\cite{Borisenko2014, Liu2014, Neupane2014a}. Recently,  a topological phase transition has been theoretically proposed in these compounds using the DFT+CPA calculations~\cite{Narayan2014} with an isovalent substitution at the cite of Bi or As by Sb or P, respectively, leading to a phase transition from the semimetalic to the semiconducting in nature with a subsequent loss of the band inversion. Thus, the system undergoes a topological phase transition from nontrivial semimetalic to trivial semiconductivity. During this phase transition, interestingly, at a critical substitution,  one would observe a novel band structure in which a lone bulk Dirac point exists in the vicinity of the Fermi level and thus  leaving out the complexity of the multiple Dirac points as a case in the parent system. In addition, such a single Dirac cone semimetal can be further tuned into a novel Weyl semimetal just by breaking one of the aforementioned two symmetries. In such a novel semimetal there exist only two Weyl nodes, contrary to the so far reported  minimal four ~\cite{Haubold2017} or higher Weyl node systems~\cite{Xu2015, Soluyanov2015, Deng2016, Tamai2016}.  Therefore, finding out such a peculiar semimetal  with the aforementioned novel band structure is of broad importance in both basic science and technology~\cite{Nayak2008}.

There exist already a couple of transport studies,  reporting on the phase transition of 3D Dirac semimetals, (Cd$_{1-x}$Zn$_x$)$_3$As$_2$ ~\cite{Sankar2015, Lu2017} and (Cd$_{1-x}$Sn$_x$)$_3$As$_2$~\cite{Sankar2015}.  However, no photoemission study on the phase transition of these systems is available till date. Here, we report the low-energy electronic structure of Cd$_3$(As$_{0.66}$P$_{0.34}$)$_2$ using the high-resolution angle-resolved photoemission spectroscopy (ARPES) and compare it with the electronic structure of the parent Cd$_3$As$_2$. Unlike in the previous transport studies~\cite{Sankar2015, Lu2017} where the Cd is replaced by Zn or Sn~\cite{Sankar2015, Lu2017}, in Cd$_3$(As$_{1-x}$P$_x$)$_2$ the As atom is replaced by the isovalent P. Our ARPES studies show gapless bulk Dirac states in the parent Cd$_3$As$_2$, while a gap of 0.23 eV is observed for the same Dirac states in the doped Cd$_3$(As$_{0.66}$P$_{0.34}$)$_2$. This observation provides a clear evidence of phase transition with the P substitution in Cd$_3$As$_2$ as predicted earlier for these systems~\cite{Narayan2014}.

% We further noticed difference in the applied photon energies between Cd$_3$As$_2$ and Cd$_3$(As$_{0.8}$P$_{0.2}$)$_2$ in order to probe the Dirac states. This confirms that with the doping the two Dirac points spread along $k_z$ get closer to form a lone Dirac cone at the $\Gamma$ point.

\begin{figure}[htbp]
	\centering
		\includegraphics[width=0.49\textwidth]{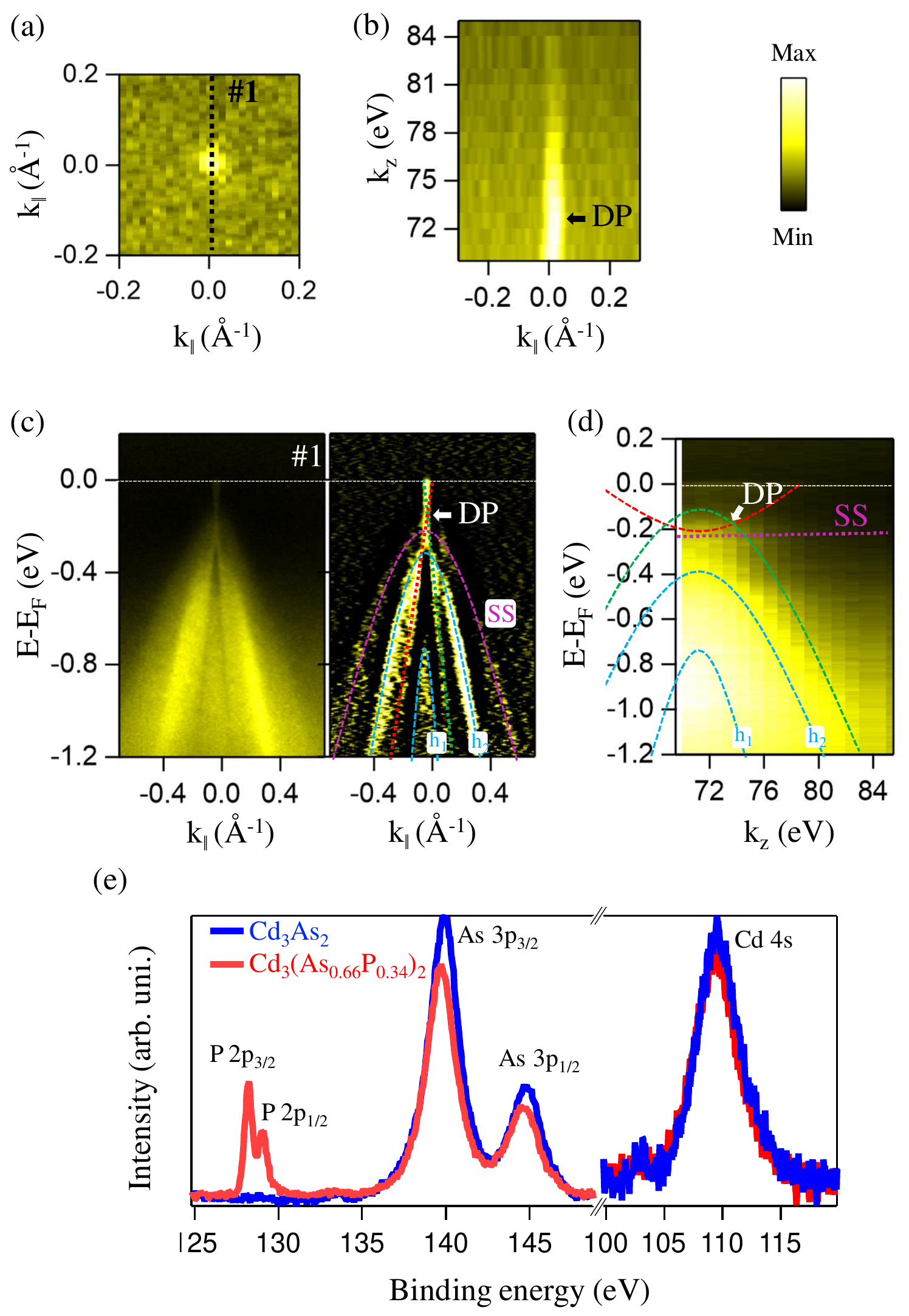}
	\caption{ARPES data of Cd$_3$As$_2$. (a) In-plane Fermi surface map. (b) $k_z$ Fermi surface map. (c) Energy distribution map (EDM) measured along the cut 1 as shown on the Fermi surface map (left panel) and second derivative of the EDM (right panel). (d) Energy distribution map taken along the $k_z$ direction.  Dirac points (DP) are located in the panels (b), (c) and (d). (e) XPS data of Cd$_3$As$_2$ and Cd$_3$(As$_{0.66}$P$_{0.34}$)$_2$. Multiplet peaks of P at the binding energy of 129 eV are confirming the element P substitution in Cd$_3$(As$_{0.66}$P$_{0.34}$)$_2$.}
	\label{1}
\end{figure}

Single crystals of Cd$_3$(As$_{1-x}$P$_{x}$)$_2$ [$x$ = 0 and 0.34(3)] were prepared by the self-flux method by mixing the elements of As and P with excess Cd in an evacuated quartz tube. The mixture then slowly cooled from 800 $^{\circ}$C to 400 $^{\circ}$C and separated the single crystals from the excess Cd flux by centrifuging the melt at 400 $^{\circ}$C~\cite{Ali2014a, Lu2017}. As obtained single crystals are structurally analyzed by the powder X-ray diffraction (XRD). The Phosphor content of the cleaved surface is determined by the energy dispersive X-ray (EDX) analysis. Single crystals of  Cd$_3$As$_2$ are identified in the orthorhombic I4$_1$/$acd$ space group with the lattice constants of $a = b = 12.6648(12) \AA$, $c = 25.516(4) \AA$ and $V = 4092.7(8) \AA^3$,   while the single crystals of Cd$_3$(As$_{0.66}$P$_{0.34}$)$_2$ are identified in the tetragonal P4$_2$/$nmc$ space group with the lattice constants of $a = b = 8.8858(4)\AA$,  $c = 12.4598(7) \AA$ and $V = 983.8(1) \AA^3$. The parameters found in the parent system are in а good agreement with available literature~\cite{Ali2014a}. Further structural details of the substituted crystals will be available elsewhere~\cite{Kataeva}. ARPES measurements were performed in BESSY II (Helmholtz Zentrum Berlin) synchrotron radiation center at the UE112-PGM2b beam-line using the "1$^3$-ARPES"~\cite{Borisenko2012a,Borisenko2012b} end station which is equipped with SCIENTA R4000 analyzer.  Depending on the incident photon energies, the total energy resolution was set between  5 and 15 meV.  The measurements were performed at a sample temperature of 1 K. Another set of ARPES measurements were performed in the Diamond light source at I05 beamline~\cite{Hoesch2017} which is equipped with SCIENTA R4000 analyzer. During the measurements at Diamond the sample temperature was at 5 K and the energy resolution was set between 10 and 20 meV depending on the incident photon energy. Difference in the band structure of these systems is not expected between 1 and 5K  as these show no phase transition within this temperature range.

\begin{figure*}[htbp]
	\centering
		\includegraphics[width=0.98\textwidth]{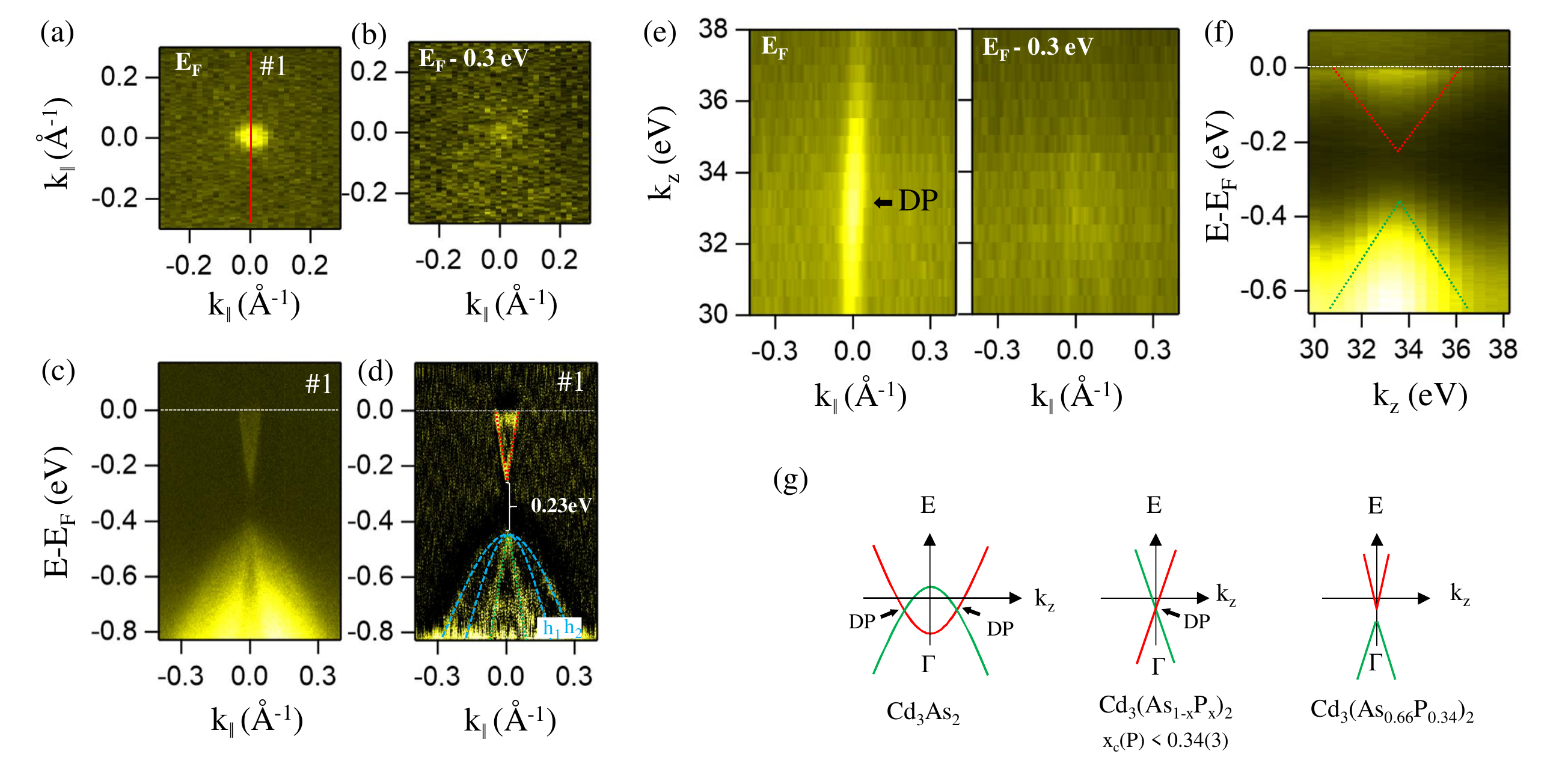}
	\caption{ARPES data of Cd$_3$(As$_{0.66}$P$_{0.34}$)$_2$. (a) In-plane Fermi surface map. (b) Constant energy map taken at 0.3 eV below $E_F$. (c) EDM taken along the cut shown on the FS map. (d) Second derivative of (c). (e) $k_z$ Fermi surface map at the Brillouin zone center (top panel) and constant energy map taken at 0.3 eV below $E_F$ (bottom panel). (f) EDM taken along the $\Gamma-Z$ direction. (g) Schematics are showing the band evolution in Cd$_3$As$_2$ with P substitution along the $k_z$ direction.}
	\label{2}
\end{figure*}

ARPES data of the parent Cd$_3$As$_2$ are shown in Figure~\ref{1}. All the data shown in Fig.~\ref{1} are measured using the $p$-polarized light. Fig.~\ref{1}(a) shows the in-plane Fermi surface (FS) map measured with a photon energy of 70 eV.  Fig.~\ref{1}(b) depicts the $k_z$ Fermi surface map measured by varying the photon energy between 70 and 85 eV in steps of 1 eV. Left panel in Fig.~\ref{1}(c) is the energy distribution map (EDM) measured along the cut 1 as shown on the FS map. Second derivative of the EDM is shown in the right panel of Fig.~\ref{1}(c). Fig.~\ref{1}(d) depicts EDM taken along the $k_z$ direction. Note that these crystals have a tendency of cleaving in the (112) plane~\cite{Ali2014a, Yi2014}. Therefore, in this manuscript the $k_z$ refers to the [112] direction. Our ARPES data of Cd$_3$As$_2$ shown in Fig.~\ref{1}  is in very good agrement with the previous ARPES reports on this system~\cite{Borisenko2014, Liu2014, Neupane2014a, Liu2014a, Yi2014}, i.e., linear dispersive gapless bulk Dirac states are observed in the vicinity of the Fermi level.  Although the band structure calculations generally predict Dirac point at the Fermi level, experimentally it has been always noticed below $E_F$ due to As vacancies~\cite{Spitzer1966}. In our data as well, we notice the Dirac point approximately at 180 meV below $E_F$. In addition to the Dirac cone, in Fig.~\ref{1}(c),  we further notice two more bulk holelike bands (h$_1$ and h$_2$) and a holelike surface state (SS) as shown by the dashed-blue and green curves, respectively. Observation of the holelike bands (h$_1$ and h$_2$) is in good agreement with previous band structure calculations on this system~\cite{Wang2013a, Ali2014a,  Borisenko2014, Yi2014, Narayan2014}.  Also, the observation of surface state (SS) is consistent with an earlier ARPES  study~\cite{Yi2014}.

\begin{figure} [tbp]
	\centering
		\includegraphics[width=0.49\textwidth]{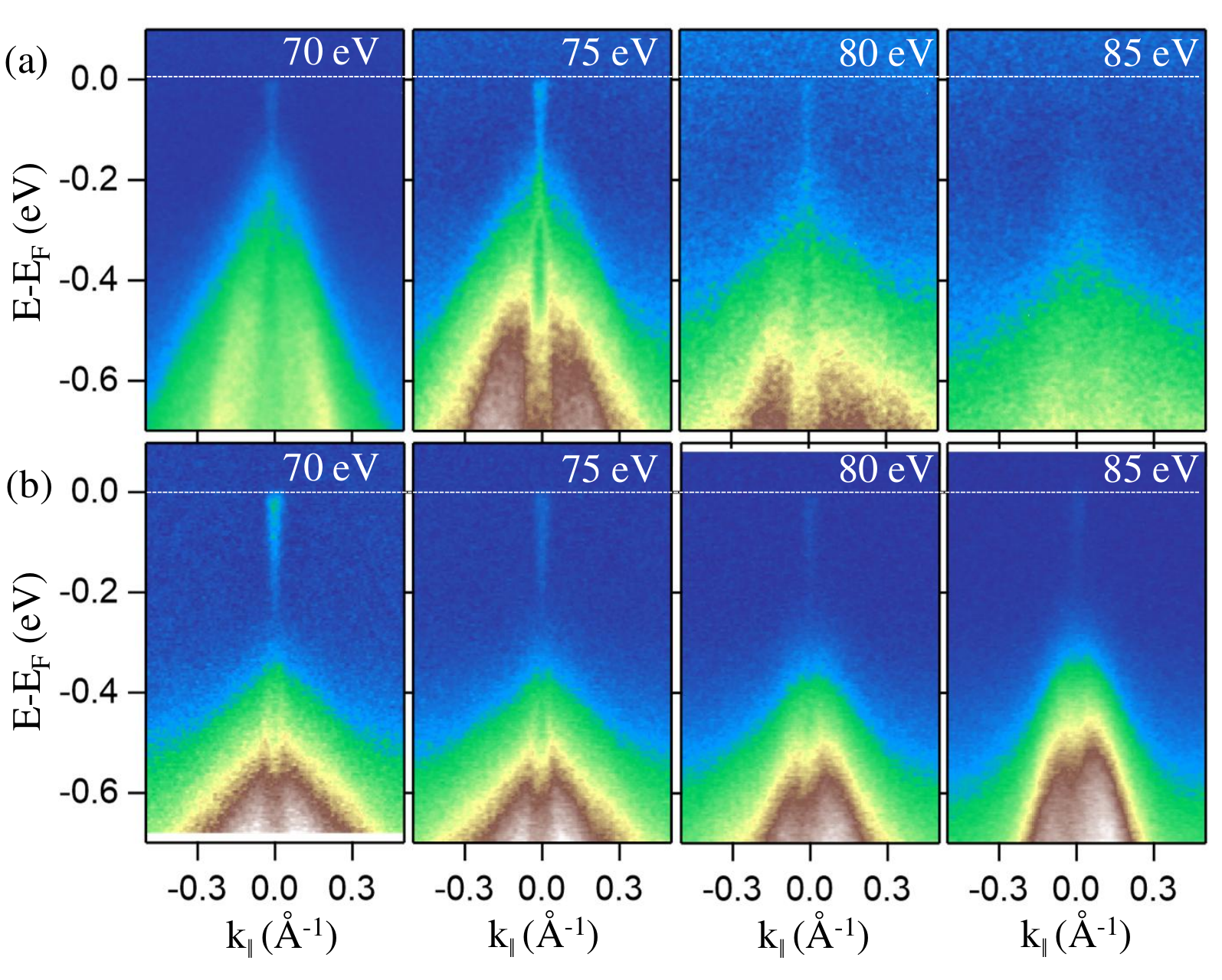}
	\caption{ Photon energy dependent EDMs from (a) Cd$_3$As$_2$ and (b) Cd$_3$(As$_{0.66}$P$_{0.34}$)$_2$.}
	\label{3}
\end{figure}

ARPES data of Cd$_3$(As$_{0.66}$P$_{0.34}$)$_2$ are shown in Figure~\ref{2}. Fig.~\ref{2}(a) depicts the in-plane Fermi surface (FS) map measured with a photon energy of 20 eV.   Fig.~\ref{2}(b) depicts the constant energy map taken at a binding energy of 0.3 eV below $E_F$.  Fig.~\ref{2}(c) is the EDM measured along the cut 1 as shown on the FS map. Fig.~\ref{2}(d) is the second derivative of ~\ref{2}(c). Left panel in Fig.~\ref{2}(e) shows the $k_z$ Fermi surface map measured by varying the photon energy between 30 and 38 eV in steps of 0.5 eV, extracting the band information around the $\Gamma$-point. Right panel in Fig.~\ref{2}(e) shows the constant energy $k_z$ map at a binding energy of 0.3 eV below $E_F$. Fig.~\ref{2}(f) shows the EDM taken along the $k_z$ direction.  Similar to the parent system, in Cd$_3$(As$_{0.66}$P$_{0.34}$)$_2$ as well, we notice two more hole pockets (h$_1$ and h$_2$) shown by the dashed-blue curves whose band tops are degenerate and coincide with the energy position of the lower half of Dirac cone. However, unlike in the parent system, no surface states have been detected in the doped compound. Further data comparison between the parent Cd$_3$As$_2$  and Cd$_3$(As$_{0.66}$P$_{0.34}$)$_2$ are shown in Figure~\ref{3}. Fig.~\ref{3} (a) shows EDMs measured with the photon energies 70, 75, 80 and 85 eV from Cd$_3$As$_2$ and Fig.~\ref{3} (b) shows similar data but measured from Cd$_3$(As$_{0.66}$P$_{0.34}$)$_2$ using the $p$-polarized light.

 As we clearly notice from Figs.~\ref{2} (c) and ~\ref{2}(d), the Dirac states are gapped out with an energy of 0.23 eV at the node due to the P substitution in Cd$_3$As$_2$. This is further confirmed from the  in-plane and out-of-plane constant energy maps taken at a binding energy of 0.3 eV [see Figs.~\ref{2} (b) and ~\ref{2}(e)] in which no spectral intensity is observed. Interestingly, the gapped out Dirac cones in Cd$_3$(As$_{0.66}$P$_{0.34}$)$_2$ are not broadened with the substitution, suggesting that the Dirac states are robust against the perturbation as suggested by the calculations~\cite{Narayan2014}.  Observation of the energy gap of 0.23 eV in Cd$_3$(As$_{0.66}$P$_{0.34}$)$_2$ is quite astonishing because earlier theoretical study on Cd$_3$(As$_{1-x}$P$_{x}$)$_2$ suggested that the non-trivial band topology persists even up to a P substitution concentration of 90\%. On the other hand, in our study we see a phase transition from a nontrivial band structure in Cd$_3$As$_2$ to a  trivial band structure in Cd$_3$(As$_{1-x}$P$_{x}$)$_2$ with a 34(3)\% of P substitution, thus suggesting that the nontrivial band topology in Cd$_3$As$_2$ survives only over a narrow P substitution range, i.e., 0$\leq$ x(P) $<$ 0.34(3). This is a bit surprising result because such a small amount of P substitution cannot result into a substantial band structure changes as the difference in the spin-orbit coupling (SOC) strength would be negligible between x(P)= 0 and x(P)=0.34(3),  which is considered to be the driving force of the phase transition of these compounds, unless in addition to the SOC there exist the other factors such as the crystal disorder. Noteworthy to mention here that the effect of crystal field splitting with the substitution of P is negligible~\cite{Dowgiallo-Plenkiewicz1978}, thus plays no role in the phase transition of this compound. In some of the previous optical measurements, contrary to the recent theoretical predictions~\cite{Narayan2014}, the critical substitution is reported to be x$_c$(P)=0.3~\cite{Radoff1972, WagnerRJ1971}. Moreover, there it is further suggested that the required substitution concentration x(P) is of 0.6 for an energy gap of 0.23 eV. Thus, our data suggest that the non-trivial topology of Cd$_3$As$_2$ is remarkably sensitive to the isovalent P substitution at the site of As. Nevertheless, the isovalent substitution of Zn at the Cd site is seemingly lesser sensitive as the non-trivial band topology in (Cd$_{1-x}$Zn$_x$)$_3$As$_2$  survives up to a substitution concentration of x$_c$(Zn)=0.38~\cite{Lu2017}. That means, in (Cd$_{1-x}$Zn$_x$)$_3$As$_2$ the gap opens at the Dirac point only at x(Zn)$>$0.38.

Moreover, with the substitution  the anisotropic band structure of Cd$_3$As$_2$ becomes isotropic in Cd$_3$(As$_{1-x}$P$_{x}$)$_2$ at the critical P concentration. That means,  instead of two Dirac points per unit cell that are spread along the $c$-axis as observed in the parent system, in Cd$_3$(As$_{1-x}$P$_{x}$)$_2$  there exists a lone Dirac cone at the $\Gamma$-point. Therefore, the anisotropic transport properties governed by the two Dirac points in the parent system~\cite{Sankar2015} may become isotropic in Cd$_3$(As$_{1-x}$P$_{x}$)$_2$ as it has one Dirac point per unit cell in all the directions and further should lead to enhancement in the charge mobility as the Dirac-Dirac point scattering is prohibited. However, so far no such systematic in-plane and out-of-plane transport studies are available on these doped systems.   Therefore, our results demand for more systematic transport studies in this direction to have a clear understanding on the advantage of having one Dirac point near the Fermi level .

In conclusion, using high-resolution angle-resolved photoemission spectroscopy we studied the low-energy electronic structure of   Cd$_3$(As$_{0.66}$P$_{0.34}$)$_2$  and compared it with the band structure of parent Cd$_3$As$_2$. Our results clearly demonstrate an energy gap of 0.23 eV  in Cd$_3$(As$_{0.66}$P$_{0.34}$)$_2$ at the Dirac point, contrast to the gapless Dirac states observed in Cd$_3$As$_2$. This observation provides a clear evidence of the phase transition in Cd$_3$As$_2$ from a topological semimetal to a trivial semiconductor with the isovalent P substitution at the As site that was predicted earlier. We further experimentally notice that the non-trivial band topology of Cd$_3$As$_2$ is remarkably sensitive to the P substitution as it survives over a narrow substitution range, i.e.,  0 $\leq$ x (P) $<$ 0.34(3) when compared to the theoretical predictions.

This work was supported under DFG Grant No. BO 1912/7-1. S.T. acknowledges support by the Department of Science and Technology, India through the INSPIRE-Faculty program (Grant No. IFA14 PH-86). I.V.M. and G.S. thank the support from RSF,Grant No.16-42-01100. We acknowledge Diamond Light Source for the time on Beamline I05 under the Proposal SI18586-1.

\bibliography{Cd3As2}

\end{document}